\documentclass[10pt,conference]{IEEEtran}
\usepackage{mathpazo}
\usepackage{times}

\usepackage{amsmath}
\usepackage{amsfonts}
\usepackage{latexsym}
\usepackage{amssymb}
\usepackage{upref}
\usepackage{theorem}
\usepackage{graphicx}
\usepackage{psfrag}
\usepackage{algorithmic}
\usepackage{algorithm}
\usepackage{verbatim}
\usepackage{cite}
\usepackage{color}

\usepackage{tikz}

\usetikzlibrary{arrows}

\theoremstyle{plain}
\theorembodyfont{\normalfont\slshape}

\newtheorem{thm}{Theorem$\!$}
\newenvironment{theorem}
{\begin{thm}\hspace*{-1ex}{\bf.}}{\end{thm}}

\newtheorem{lem}{Lemma$\!$}

\newtheorem{prop}{Proposition$\!$}

\newtheorem{cor}{Corollary$\!$}

\newtheorem{defn}{Definition$\!$}

\newtheorem{xmpl}{Example$\!$}
\newenvironment{example}{\begin{xmpl}\hspace*{-1ex}{\bf.}}{\end{xmpl}}

\newtheorem{cnstr}{Construction$\!$}
\newenvironment{construction}{\begin{cnstr}\hspace*{-1ex}{\bf.}}{\end{cnstr}}

\newtheorem{asmp}{Assumption$\!$}

\newcounter{enumrom}
\renewcommand{\theenumrom}{(\roman{enumrom})}

\makeatletter
\renewcommand{\@endtheorem}{\endtrivlist}
\makeatother

\makeatletter
\renewcommand{\thefigure}{{\@arabic\c@figure}}
\renewcommand{\fnum@figure}{{\bf Figure\,\thefigure}}
\makeatother

\newcommand{\be}[1]{\begin{equation}\label{#1}}
\newcommand{\ee}{\end{equation}}

\renewcommand{\le}{\leqslant}

\newcommand{\Cref}[1]{Co\-ro\-lla\-ry\,\ref{#1}}

\DeclareMathAlphabet{\mathbfsl}{OT1}{cmr}{bx}{it}

\outer\def\proclaim #1. #2\par{\medbreak
 \noindent{\bf#1.\enspace}{\sl#2\par}%
 \ifdim\lastskip<\medskipamount \removelastskip\penalty55\medskip\fi}

\mathchardef\inn="3232
\renewcommand{\in}{{\,\inn\,}}

\allowdisplaybreaks[1]

\begin{document}

\IEEEoverridecommandlockouts 
\title{\Huge\bf  Repair-Optimal MDS Array Codes over GF(2)}


\author{
\IEEEauthorblockN{\textbf{Eyal En Gad$^{\ast}$, Robert Mateescu$^{\dagger}$, Filip Blagojevic$^{\dagger}$ , Cyril Guyot$^{\dagger}$} and \textbf{Zvonimir Bandic$^{\dagger}$}}
\IEEEauthorblockA{  $^{\ast}$Electrical  Engineering, California Institute of Technology, Pasadena, CA 91125. \\
$^{\dagger}$HGST Research, San Jose, CA 95135. \\
{\it $^{\ast}$eengad@caltech.edu, $^{\dagger}$\{robert.mateescu,zvonimir.bandic\}@hgst.com}\vspace*{-4.0ex}}

}

\maketitle

\begin{abstract}
Maximum-distance separable (MDS) array codes with high rate and an optimal repair property were introduced recently. These codes could be applied in distributed storage systems, where they minimize the communication and disk access required for the recovery of failed nodes. However, the encoding and decoding algorithms of the proposed codes use arithmetic over finite fields of order greater than $2$, which could result in a complex implementation.

In this work, we present a construction of $2$-parity MDS array codes, that allow for optimal repair of a failed information node using XOR operations only. The reduction of the field order is achieved by allowing more parity bits to be updated when a single information bit is being changed by the user.
\end{abstract}

\section{Introduction}
\label{sec:intro}

MDS array codes are highly applicable in modern data storage systems. Array
codes are non-binary erasure codes, where each symbol is a column of elements
in a two dimensional array, and is stored on a different storage node in the
system.  In traditional erasure codes, the decoder uses all of the available
codeword symbols for the recovery of erased symbols. However, in distributed
storage systems, this property requires the transmission of an entire array
over the network for the recovery of failed nodes. And since node failures
are common, the network load caused by node recovery became a major
constraint to the application of erasure codes in such systems \cite{KhaBurPlaHua11}.
%

For that reason, a lot of attention has been drawn recently to the
minimization of the communication required for node recovery. The total amount of information communicated in the network during recovery is called the \emph{repair bandwidth} \cite{DimGodWuWaiRam10}.
In this work we focus on the practical case of systematic MDS array codes with $2$ parity nodes. In this case, when $2$ nodes are erased, the entire information must be transmitted in order the repair the erased nodes. However, when only a single node is erased, the required repair bandwidth can be lower. It was shown in \cite{DimGodWuWaiRam10} that the repair bandwidth must be at least $1/2$ of the entire available information in the array.
Subsequently, several constructions were designed to achieve that lower bound \cite{CadJafMal10,TamWanBru11,PapDimCad11,CadHuaLi11}.

Beside the repair bandwidth, another important parameter of array codes is the \emph{update measure}. In systematic array codes, the elements of the information nodes are called information elements, and those in the parity nodes are called parity elements. The update measure is defined as the number of parity elements that need to change each time an information element is being updated. For MDS array
codes, the update measure cannot be smaller than the number of parity nodes.
For the codes in \cite{CadJafMal10,TamWanBru11,PapDimCad11,CadHuaLi11}, the update measure is
optimal. Another property of these codes is that the elements of the nodes belong to a finite field of order at least $3$. This property can make the codes difficult for hardware implementation. However, it was shown in these papers that for MDS codes with optimal repair bandwidth and optimal update measure, the node elements cannot belong to GF($2$). This is the
point of departure of this work. Instead of designing codes with optimal
update measure, we focus on the design of codes with node elements in GF($2$), with the price of a higher update measure. This offers a different trade-off, that can find a wide array of applications.

The main contribution of this work is a construction of systematic MDS array codes with node elements in GF($2$). The construction have a similar structure to the ones described in \cite{CadJafMal10,TamWanBru11,PapDimCad11,CadHuaLi11}. The codes have $2$ parity nodes, and a failure of any information node can be recovered with the access to only $1/2$ of the available information in the array. Note that in general, the amount of \emph{accessed} information in node recovery can be different from the repair bandwidth. Specifically, the total access can be higher than the total bandwidth, but not lower, since there is no reason to communicate more than what is accessed. For that reason, our construction have both optimal access and optimal repair bandwidth in the case of a single information node failure. However, in the case of a parity node failure, the entire information array needs to be transmitted for the recovery. But this is not a major drawback, since a parity node failure does not reduce the availability of the stored data to the user, and thus its
recovery can be done offline, and does not affect the system performance significantly.
%
The update measure in
our construction is different for different elements. For $k$ information
nodes, where $k$ is odd, the expected update is $1/2\cdot\lfloor k/2\rfloor\
+2$, and the worst-case update is $\lfloor k/2\rfloor\ +2$.

The rest of the paper is organized as following: In section \ref{sec:dem} we demonstrate the key principles of the construction by simple examples. Next, the construction is described formally in section \ref{sec:construction}, with an additional example. Lastly, the properties of the constructions are proven in section \ref{sec:properties}, and concluding remarks are brought in section \ref{sec:conclusions}.

\section{Demonstrating Examples}
\label{sec:dem}

\begin{figure}[t]
\tikzstyle{spec}= [rectangle,draw,text centered,text width=18pt, text
height=3pt,text depth=1pt, fill=black!20]
\tikzstyle{norm} = [rectangle,draw,text centered,text width=18pt, text
height=3pt,text depth=1pt,]

\centering
\vspace{10pt}
\begin{tikzpicture}[yscale=.5, xscale=2.5, node distance=0.5cm, auto]
          \node[norm, name=c-0-0] at (0,0) {$a_{0,0}$};
          \node[spec, name=c-1-0] at (-1,0) {$a_{0,1}$};
          \node[norm, name=c-0-1] at (0,-1) {$a_{1,0}$};
          \node[norm, name=c-1-1] at (-1,-1) {$a_{1,1}$};

          \path (c-0-0.west) edge[-] (c-1-0.east);
          \path (c-0-0.west) edge[-] (c-1-1.east);
          \path (c-0-1.west) edge[-] (c-1-0.east);
          \path (c-0-1.west) edge[-] (c-1-1.east);

\end{tikzpicture}

\[
\begin{array}{|c|c||l|l|}
  \hline
  a_{0,1} & a_{0,0} & h_0 = a_{0,1} + a_{0,0} &  b_0 = a_{0,0} + a_{1,1} \\ \hline
  a_{1,1} & a_{1,0} & h_1 = a_{1,1} + a_{1,0} &  b_1 = a_{1,0} + a_{0,1} + a_{0,0} \\ \hline
\end{array}
\]
\caption{Decoding a butterfly cycle.}
\label{fig:2nodes}
\vspace{-10pt}

\end{figure}

A basic principle of the construction can be demonstrated in the case of two information nodes, shown in  Figure~\ref{fig:2nodes}. In this case, each of the columns contains two
elements. The information element in row $i$ and column $j$ is denoted as $a_{i,j}$. As is the case in the rest of the paper, there are two parity nodes. The first parity node is called the horizontal node, as its elements are encoded as the horizontal parities. The horizontal element in row $i$ is denoted as $h_i$, and its value is the parity of the information elements in row $i$. The summations in the table of Figure~\ref{fig:2nodes} are taken modulo $2$ without mention, as are all of the summations of bits in the rest of the paper.

The second parity node is called the butterfly node, and its element in row $i$ is denoted as $b_i$. The reason for the name will be clear in the next example. In the figure above the table, the horizontal elements correspond to the horizontal lines, and the butterfly elements to the diagonal lines. However, as shown in the table, the encoding of $b_1$ contains also the element $a_{0,0}$. In the figure, this is symbolized by the dark color of $a_{0,1}$, that signifies that the element to its right is also added to the corresponding butterfly element.

Now consider the case that column $1$ is erased. In this case the column can be decoded using the available elements of row $0$ only, by setting $a_{0,1}=h_0+a_{0,0}$, and $a_{1,1}=b_0+a_{0,0}$. Since the decoder accesses only half of the elements in each available column, and only half of the available information in total, we say that the \emph{access ratio} is $1/2$.


Since we claim that the code is MDS, consider the case that both information nodes are erased.
Notice that if $a_{0,0}$ was not included in $b_1$, the code could not recover from the loss of the two information nodes. However,
the addition of $a_{0,0}$ to $b_1$, which corresponds to the dark element
$a_{0,1}$, allows to break the cycle, and create a decoding chain. From $h_0
+ b_1$ we obtain $a_{1,0}$. In the decoding chain that remains in Figure~\ref{fig:2nodes}
if we eliminate the diagonal $a_{0,1}, a_{1,0}$, we now have all the segments
and the end element, and therefore all the other three elements can be
decoded.
Notice that the addition of $a_{0,0}$ also increases the update measure. If the user wants to change the value of $a_{0,0}$, the encoder needs to update the element $b_1$, in addition to $h_0$ and $b_0$.
The code in Figure~\ref{fig:2nodes} is also the simplest version of
the EVENODD code~\cite{BlaBraBruMen95}.

Now consider the case of $3$
information nodes. In this case, the construction requires that the nodes
contain $4$ elements, where in general, for $k$ information nodes, the number of
elements is $2^{k-1}$. Although the size of the column is exponential in the
number of columns, this is still practical because the usual number of
storage nodes is typically between 10 and 20, and the element of a column is
a single bit.

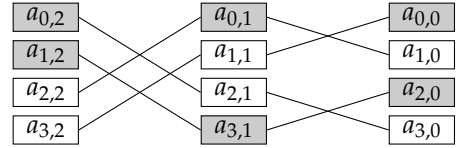
\begin{figure}[t]
\tikzstyle{spec}= [rectangle,draw,text centered,text width=18pt, text
height=3pt,text depth=1pt, fill=black!20]
\tikzstyle{norm} = [rectangle,draw,text centered,text width=18pt, text
height=3pt,text depth=1pt,]

\centering
\vspace{10pt}
\begin{tikzpicture}[yscale=.5, xscale=2.5, node distance=0.5cm, auto]
     \foreach \y in {0,2}
          \node[spec, name=c-0-\y] at (0,-\y) {$a_{\y,0}$};
     \foreach \y in {1,3}
          \node[norm, name=c-0-\y] at (0,-\y) {$a_{\y,0}$};

     \foreach \y in {0,3}
          \node[spec, name=c-1-\y] at (-1,-\y) {$a_{\y,1}$};
     \foreach \y in {1,2}
          \node[norm, name=c-1-\y] at (-1,-\y) {$a_{\y,1}$};

     \foreach \y in {0,1}
          \node[spec, name=c-2-\y] at (-2,-\y) {$a_{\y,2}$};
     \foreach \y in {2,3}
          \node[norm, name=c-2-\y] at (-2,-\y) {$a_{\y,2}$};

    \foreach \a / \b in {0/1,2/3}
    {
          \path (c-0-\a.west) edge[-] (c-1-\b.east);
          \path (c-0-\b.west) edge[-] (c-1-\a.east);
    }
    \foreach \a / \b in {0/2,1/3}
    {
          \path (c-1-\a.west) edge[-] (c-2-\b.east);
          \path (c-1-\b.west) edge[-] (c-2-\a.east);
    }

\end{tikzpicture}
\caption{The encoding of the butterfly node.}
\label{fig:3nodes}
\vspace{-10pt}
\end{figure}

The horizontal elements are encoded in the same way as before, as the parity of the rows. The butterfly node is now encoded with correspondence to its name, where each $b_i$ is encoded according to the line in the \emph{butterfly diagram} of Figure~\ref{fig:3nodes} that starts at element
$a_{i,0}$. Note that we draw the butterfly with column $0$ on the right side.
The element $b_i$ is encoded as the parity of the elements in this line, and
in addition, if there are \emph{dark} elements in the line, according to
Figure~\ref{fig:3nodes}, extra elements are added to $b_i$. For each dark
element in the line, the element to its right (cyclicly) is also added to
$b_i$. In the general case of $k$ information nodes, the $\lfloor k/2
\rfloor$ elements to the right of a dark elements are added (for odd $k$, see
details in section~\ref{sec:construction}). The careful addition of extra elements in the butterfly
parity, corresponding to the dark elements, is what allows the
computation to be done in GF(2).
 In this example,
$b_0=a_{0,0}+a_{1,1}+a_{3,2}+a_{0,2}$. The elements $a_{0,0}$, $a_{1,1}$,
$a_{3,2}$ come form the butterfly line; additionally, since $a_{0,0}$ is dark, the
element to its right (cyclicly), $a_{0,2}$, is also added. Similarly, $b_{2}
= a_{2,0}+a_{3,1}+a_{1,2}+a_{2,2}+a_{3,0}+a_{1,1}$.

The dark elements in Figure~\ref{fig:3nodes} are those $a_{i,j}$ for which the $j$-th bit in the binary
representation of $i$ over $k-1$ bits, is equal to the $(j-1)$-th bit, where
the $-1$-th bit is considered as $0$. For example, $a_{0,1}$ is dark since
the bit $1$ of $0$ is equal to bit $0$ of $0$. Now consider the case that
node $1$ is failing and needs to be reconstructed. The decoding method for a single node failure is simple: recover the dark
elements by the horizontal parities, and the white elements by the butterfly
parity. In the example, we set $a_{0,1}=h_0+a_{0,0}+a_{0,2}$ and
$a_{3,1}=h_3+a_{3,0}+a_{3,2}$, and the dark elements are recovered. For the
white elements, we set $a_{1,1}=b_0+a_{0,0}+a_{3,2}+a_{0,2}$ and
$a_{2,1}=b_3+a_{3,0}+a_{0,2}+a_{0,1}$ (where $a_{0,1}$ was recovered by the
horizontal parity). Notice that according to this method, the decoder only access rows $0$ and $3$, and the access ratio is $1/2$.

Now consider the case that nodes $0$ and $1$ fail. We can see in Figure~\ref{fig:3nodes} that there are two decoding cycles, the cycle of rows $0$ and $1$, and that of rows $2$ and $3$. For this decoding, we can ignore the fact the $a_{0,0}$ and $a_{2,0}$ are dark, since the added elements are in column $2$, which is available. Therefore, the top cycle becomes identical to the previous example, and can be decoded in the same way. Note that the bottom cycle could not be decoded before the top one. That is since the dark elements of column $2$ imply that $a_{0,1}$ and $a_{1,1}$ are added to the butterfly lines of the bottom cycle, and since they are unknown, the cycle cannot be decoded. However, after the decoding of the top cycle, the bottom cycle can be decoded in the same way. In the case of more information columns, the order needed for the decoding of the cycles is related to a binary reflected Gray code, and is described in the next section.

\section{Code Construction}
\label{sec:construction}

For the presentation of the construction we use extra notation. Let $[n]=\{0,1,\dots,n-1\}$.
For integers $i$ and $j$, $i\oplus j$ denotes the bitwise XOR operation between the binary representations of $i$ and $j$, where the result is converted back to be an integer.
The expression $i(j)$ denotes the $j$-th bit in the binary representation of $i$, where $i(0)$ is the least significant bit. If $j$ is negative, $i(j)$ is defined to be $0$.
The construction requires that $k$ is odd. If the number of information nodes is even, assume that there is an extra node, where the values of all its entries are $0$. The construction is now described formally.

\begin{construction}
\label{con:butterfly}
For each pair $(i,j)\in[2^{k-1}]\times[k]$, define a set $B_{i,j}$ as following. If $i(j)\ne i(j-1)$, let $B_{i,j}=\{(i,j)\}$. Else, let
\[B_{i,j}=\{(i,j'):j-j'\le\lfloor k/2\rfloor \text{ (mod $k$)}\}.\]
 Next, let $\ell_{i,j}=i\oplus (2^j-1)$, and for each $i\in [2^{k-1}]$, define a set \[B_i=\cup_{j\in[k]}B_{\ell_{i,j},j}.\]

\textbf{Encoding}:
For each $i\in[2^{k-1}]$, set
\[h_i=\sum_{j\in[k]}a_{i,j}, \qquad b_i=\sum_{(i',j')\in B_i}a_{i',j'}.\]

\textbf{Single failure decoding}:
If the failed node is a parity node, use the encoding method. If information node $j$ failed, for each $i\in[2^{k-1}]$ recover $a_{i,j}$ as following: If $i(j-1)=i(j)$, set $a_{i,j}=h_i+\sum_{j'\ne j}a_{i,j'}$. Else, set
\[a_{i,j}=b_{\ell_{i,j}}+\sum_{(i',j')\in B_{\ell_{i,j}}\setminus\{(i,j)\}}a_{i',j'}.\]

\textbf{Double failure decoding}:
If both failed nodes are parity nodes, use the encoding method. If one of them is the butterfly node and the other is the information node $j$, then for each $i\in[2^{k-1}]$, set $a_{i,j}=h_i+\sum_{j'\ne j}a_{i,j'}$, and then encode the butterfly node.

If the horizontal node fails together with the information node $j$, decode as following: For $i=0,1,\dots,2^{k-1}-1$, find $i'$ according to Algorithm \ref{alg:find1}, and set

\[a_{i',j}=b_{\ell_{i',j}}+\sum_{(i'',j'')\in B_{\ell_{i',j}}\setminus\{(i',j)\}}a_{i'',j''}.\label{eq:i'j}\]

After node $j$ is decoded, encode the horizontal node.

Finally, if two information nodes failed, denote their indices as $j_0,j_1$, such that $j_1-j_0\le\lfloor k/2\rfloor\text{ (mod $k$)}$.
Next, for $i=0,1,\dots,2^{k-2}-1$, find $i_0,i_1$ according to Algorithm \ref{alg:find2}, and set, sequentially,

{\allowdisplaybreaks
\begin{align}
a_{i_1,j_0}=&h_{i_0}+\sum_{j'\in[k]\setminus\{j_0,j_1\}}a_{i_0,j'}\nonumber\\
&+b_{\ell_{i_1,j_0}}+\sum_{(i',j')\in B_{\ell_{i_1,j_0}}\setminus\{(i_1,j_0),(i_0,j_0),(i_0,j_1)\}}a_{i',j'}\label{eq:i1j0}\\
a_{i_1,j_1}=&h_{i_1}+\sum_{j'\in[k]\setminus\{j_1\}}a_{i_1,j'}\label{eq:i1j1}\\
a_{i_0,j_0}=&b_{\ell_{i_0,j_0}}+\sum_{(i',j')\in B_{\ell_{i_0,j_0}}\setminus\{(i_0,j_0),(i_1,j_1)\}}a_{i',j'}\label{eq:i0j0}\\
a_{i_0,j_1}=&h_{i_0}+\sum_{j'\in[k]\setminus\{j_1\}}a_{i_0,j'}\label{eq:i0j1}
\end{align}
}
\end{construction}

\begin{algorithm}
\caption{Find $i'$. }
\label{alg:find1}
\begin{algorithmic}[1]
\STATE Inputs: $i\in[2^{k-1}]$
\STATE Output: $i'\in[2^{k-1}]$
\STATE $i'(k-1)\leftarrow 0$
\FOR{$j'=k-2$ \TO $j'=j$}
\STATE $i'(j')\leftarrow i'(j'+1) + i(j'-1)$
\ENDFOR
\FOR{$j'=0$ \TO $j'=j-1$}
\STATE $i'(j')\leftarrow i'(j'-1) + i(j')$
\ENDFOR
\end{algorithmic}
\end{algorithm}

\begin{algorithm}
\caption{Find $i_m$. }
\label{alg:find2}
\begin{algorithmic}[1]
\STATE Inputs: $m\in\{0,1\},i\in[2^{k-2}]$
\STATE Output: $i_m\in[2^{k-1}]$
\STATE $i_m(k-1)\leftarrow 0$
\STATE $s\leftarrow\arg\max_{i'\in\{0,1\}}\{j_{i'}\}$
\FOR{$j=k-2$ \TO $j=j_s$}
\STATE $i_m(j)\leftarrow i_m(j+1) + i(j-1)$
\ENDFOR
\FOR{$j=0$ \TO $j=j_{1-s}-1$}
\STATE $i_m(j)\leftarrow i_m(j-1) + i(j)$
\ENDFOR
\STATE $i_m(j_1-s)\leftarrow i_m(j_1+s-1)+m$
\IF{$j_s-j_{1-s}>1$}
\FOR {$j=j_1+1-3s$ \TO $j_0+s-1$}
\STATE $i_m(j)\leftarrow i_m(j+2s-1) + i(j+s-1)$
\ENDFOR
\ENDIF
\end{algorithmic}
\end{algorithm}

\begin{figure}[t]
\tikzstyle{spec}= [rectangle,draw,text centered,text width=18pt, text
height=3pt,text depth=1pt, fill=black!20]
\tikzstyle{norm} = [rectangle,draw,text centered,text width=18pt, text
height=3pt,text depth=1pt,]

\centering
\vspace{10pt}
\begin{tikzpicture}[yscale=.5, xscale=2.5, node distance=0.5cm, auto]
     \foreach \y in {0,2,4,6}
          \node[spec, name=c-0-\y] at (0,-\y) {$a_{\y,0}$};
     \foreach \y in {1,3,5,7}
          \node[norm, name=c-0-\y] at (0,-\y) {$a_{\y,0}$};

     \foreach \y in {0,3,4,7}
          \node[spec, name=c-1-\y] at (-1,-\y) {$a_{\y,1}$};
     \foreach \y in {1,2,5,6}
          \node[norm, name=c-1-\y] at (-1,-\y) {$a_{\y,1}$};

     \foreach \y in {0,1,6,7}
          \node[spec, name=c-2-\y] at (-2,-\y) {$a_{\y,2}$};
     \foreach \y in {2,3,4,5}
          \node[norm, name=c-2-\y] at (-2,-\y) {$a_{\y,2}$};

     \foreach \y in {0,1,2,3}
          \node[spec, name=c-3-\y] at (-3,-\y) {$a_{\y,3}$};
     \foreach \y in {4,5,6,7}
          \node[norm, name=c-3-\y] at (-3,-\y) {$a_{\y,3}$};

    \foreach \a / \b in {0/1,2/3,4/5,6/7}
    {
          \path (c-0-\a.west) edge[-] (c-1-\b.east);
          \path (c-0-\b.west) edge[-] (c-1-\a.east);
    }
    \foreach \a / \b in {0/2,1/3,4/6,5/7}
    {
          \path (c-1-\a.west) edge[-] (c-2-\b.east);
          \path (c-1-\b.west) edge[-] (c-2-\a.east);
    }
    \foreach \a / \b in {0/4,1/5,2/6,3/7}
    {
          \path (c-2-\a.west) edge[-] (c-3-\b.east);
          \path (c-2-\b.west) edge[-] (c-3-\a.east);
    }

\end{tikzpicture}
\caption{The butterfly construction with $4$ information nodes}
\label{fig:4nodes}
\vspace{-10pt}
\end{figure}
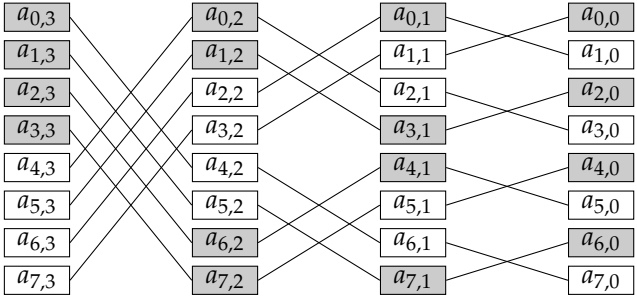

\begin{example}
Let the number of information nodes be $4$, and therefore $k=5$ (the next odd number). Now let $j_0=0$, $j_1=2$ be the failed nodes. Notice that $j_1-j_0=2\le\lfloor k/2 \rfloor=2$, as required. Assume that iterations $0$ and $1$ of the decoding were already performed, and now $i=2=10_2$. To find $i_0$ and $i_1$, we perform Algorithm \ref{alg:find2}. By lines $3$ and $4$, $i_0(4)=0$, and $s=1$. In lines $5-7$, we first set $j=3$, and $i_0(3)=0+i(2)=0+0=0$. Next, $i_0(2)=0+1=1$, and the loop is finished. Since $j_0=0$, we skip the loop in lines $8$-$10$. In line $11$, we set $i_0(1)=1+0=1$. In lines $13-15$, $j=0$, and $i_0(0)=i_0(1)+i(0)=1+0=1$, and in conclusion, $i_0=111_2=7$. It can be verified that Algorithm \ref{alg:find2} always sets $i_1=i_0\oplus(2^{j_1}-1)\oplus(2^{j_0}-1)$, and in this case, $i_1=i_0\oplus 11_2=100_2=4$.

Algorithm \ref{alg:find2} can be interpreted visually by observing Figure \ref{fig:4nodes}. In each row, consider the dark elements as $'0'$s, and the white elements as $'1'$s, and ignore columns $j_0$ and $j_1$. The rows $i_0$ and $i_1$ are the ones for which the binary number resulting from this observation is equal to $i$. In the current example, both rows $4$ and $7$ have a dark element in column $1$, and a white element in column $3$, corresponding to $i=10_2$. Note that $a_{4,2}$ and $a_{7,0}$ are in the same butterfly line, and the same is true for $a_{4,0}$ and $a_{7,2}$. The butterfly line of $a_{4,0}$ contains the white element $a_{5,1}$ and the dark element $a_{3,3}$, which are both available. Since $a_{3,3}$ is dark, the lost element $a_{3,2}$ and available element $a_{3,1}$ are also included in the same parity. So successful decoding could be made only if row $3$ was already decoded in a previous iteration. But it is easy to find by observation the iteration on which row $3$ is decoded. Since both $a_{3,1}$ and $a_{3,3}$ are dark, row $3$ was decoded in iteration $00_2=0$ which is earlier than $i=2$. Applying the same argument for the line of $a_{7,0}$, we could see that all of its elements in rows other than $4$ and $7$ are available.

To start the decoding, Algorithm \ref{alg:find2} chooses $i_0$ as the row with a dark element in column $j_1$, such that $a_{i_1,j_0}$ could be the first decoded element. In this example, $i_0=7$, and indeed $a_{7,2}$ is a dark element. Therefore, the decoding of rows $4$ and $7$ could follow directly as described by Equations \eqref{eq:i1j0}-\eqref{eq:i0j1}. We also note that Algorithm \ref{alg:find1} have the same visual interpretation.
\end{example}

\section{Code Properties}
\label{sec:properties}

In this section we show that the codes have optimal access, that they are MDS, and present their update measure. The first Theorem proves that the single failure decoding function of Construction \ref{con:butterfly} accesses only half of the elements in each surviving node, and thus Construction \ref{con:butterfly} is said to be "repair-optimal".

\begin{theorem}
\label{th:repair_optimal}
\textbf{Optimal Repair:} The single failure decoding function of Construction \ref{con:butterfly} decodes any failed information node correctly, and it accesses only $1/2$ of the elements in each of the surviving nodes.
\end{theorem}

\begin{IEEEproof}
Let $j$ be the failed node.
First, note that the fraction of elements $i\in[2^{k-1}]$ s.t. $i(j-1)=i(j)$ is $1/2$, and therefore the decoder accesses half of the elements in the horizontal node. Next, note that when $j$ is fixed, the function $f(i)=i\oplus (2^{j}-1)$ is a permutation, and therefore the decoder also access only half of the elements in the butterfly node. Finally, we will show that for each accessed element $a_{i',j'}$ in the information nodes, $i'(j-1)=i'(j)$, and thus the decoder only access half of the elements in each node, and the repair ratio is $1/2$.

Let $a_{i,j}$ be a decoded element, and $a_{i',j'}$ be an element that is accessed in the decoding process. If $i(j-1)=i(j)$, then $i'=i$ by the decoding function, and thus $i'(j-1)=i'(j)$. Else, note that by the encoding process, $i'=i\oplus(2^j-1)\oplus(2^{j''}-1)$, for some $j''\ne j$. If $j''>j$, then $i'(j-1)=i(j-1)$, and $i'(j)=i(j)+1$. And if $j''<j$, then $i'(j-1)=i(j-1)+1$, and $i'(j)=i(j)$. In both cases,
 \[i'(j-1)+i'(j)=i(j-1)+i(j)+1=1+1=0,\]
  and therefore $i'(j)=i'(j-1)$, and the proof is completed.
\end{IEEEproof}

The next Theorem verify the MDS property of the Construction.

\begin{theorem}
\textbf{MDS:} The double failure decoding function of Construction \ref{con:butterfly} decodes the failure of any two nodes correctly.
\end{theorem}

\begin{IEEEproof}

In the case that one of the failed nodes is the butterfly node, the proof is trivial by the the encoding method. If the horizontal node failed together with the information node $j$, we need to show that in each iteration $i$, all of the elements $a_{i'',j}$, where $(i'',j)\in B_{\ell_{i',j}}\setminus\{(i',j)\}$, were decoded in a previous iteration. To prove that, note that by the definition of the set, if $(i'',j)$ is in  $B_{\ell_{i',j}}\setminus\{(i',j)\}$, than there exists $j'$ such that $i''=i'\oplus(2^j-1)\oplus(2^{j'}-1)$ and $i''(j')=i''(j'-1)$. So it is enough to show that for each $j'\ne j$, such that $i''(j')=i''(j'-1)$, $a_{i'',j}$ was decoded in a previous iteration.

We prove this by induction on the iteration $i$. In the base case, $i=0$. and according to Algorithm \ref{alg:find1}, $i'=0$ as well. Now by the definition of $i''$, $i''(j')\ne i''(j')$, and the base case is proven.

For the induction step, assume that $i''(j')=i''(j'-1)$. By the definition of $i''$, $i'(j')+i'(j'-1)=i''(j')+i''(j'-1)+1=1$. In addition, for any $j''\ne j'$, $i'(j'')+i'(j''-1)=i''(j'')+i''(j''-1)$. Therefore, according to Algorithm \ref{alg:find1}, the iteration in which $a_{i'',j}$ needs to be decoded differ from $i$ in exactly one bit. And since $i'(j')\ne i'(j'-1)$, the value of that bit in $i$ is $1$, and therefore $i$ is a later iteration, and $a_{i'',j}$ was decoded before. So by the induction hypothesis, $a_{i'',j}$ is known, and the induction is proven. So, by the encoding of the butterfly elements, column $j$ is decoded successfully, and the horizontal node can be encoded afterwards.

In the case that both failed nodes are information nodes, the proof is very similar. First we need to show that all of the terms in Equation \eqref{eq:i1j0} are known when $a_{i_1,j_0}$ is being decoded. For each $j'\in[k]\setminus\{j_0,j_1\}$, $a_{i_0,j'}$ is known since it's an element of a surviving node. For $a_{i'',j''}$ where $(i'',j'')\in B_{\ell_{i_1,j_0}}\setminus\{(i_1,j_0),(i_0,j_0),(i_0,j_1)\}$, we use induction on $i$ again.

First, notice that $(i_0,j_1)$ is actually in $B_{\ell_{i_1,j_0}}$. That is since according to Algorithm \ref{alg:find2}, $i_0=i_1\oplus(2^{j_0}-1)\oplus(2^{j_1}-1)$, where the difference between $i_0$ and $i_1$ comes form lines $11-16$ in the Algorithm. Therefore, $\ell_{i_1,j_0}=i_1\oplus(2^{j_0}-1)=i_0\oplus(2^{j_1}-1)=\ell_{i_0,j_1}$. In addition, by line $11$ of Algorithm \ref{alg:find2}, $i_0(j_1)=i_0(j_1-1)$, and therefore, $B_{i_0,j_1}=\{(i_0,j':j_1-j'\le\lfloor k/2\rfloor \text{ (mod $k$)}\}$, which implies that $(i_0,j_0)\in B_{\ell_{i_1,j_0}}$ as well.
The inductive argument follows the same lines as in the previous case, and is therefore omitted.

%

At this point, we know that all of the terms in Equation \ref{eq:i1j0} are known. Now notice that $h_{i_0}+\sum_{j'\in[k]\setminus\{j_0,j_1\}}a_{i_0,j'}=a_{i_0,j_0}+a_{i_0,j_1}$, and $b_{\ell_{i_1,j_0}}+\sum_{(i',j')\in B_{\ell_{i_1,j_0}}\setminus\{(i_1,j_0),(i_0,j_0),(i_0,j_1)\}}a_{i',j'}=a_{i_1,j_0}+a_{i_0,j_0}
+a_{i_0,j_1}$, and therefore, $a_{i_1,j_0}$ is decoded correctly. After $a_{i_1,j_0}$ is decoded, it can be seen directly by Equation \eqref{eq:i1j1} that $a_{i_1,j_1}$ can be decoded correctly as well. As for $a_{i_0,j_0}$, it can be shown by the same argument that we used for $a_{i_1,j_0}$, that it could be decoded successfully. And finally, the decoding of $a_{i_0,j_1}$ also follows immediately.
\end{IEEEproof}

Lastly, we present the update measure of the Construction.

\begin{theorem}
\textbf{Update:} The expected update measure of Construction \ref{con:butterfly} is $1/2\cdot\lfloor k/2\rfloor +2$, and the worst-case is $\lfloor k/2\rfloor+2$.
\end{theorem}

\begin{IEEEproof}
For a uniformly-distributed randomly-picked pair $(i,j)\in[2^{k-1}]\times[k]$, the probability that $i(j)=i(j-1)$ is $1/2$. Therefore, in addition to $B_{\ell_{i,j},j}$, the expected number of sets $B_{i',j'}$ that contain $(i,j)$ is $1/2\cdot\lfloor k/2\rfloor+1$. In the case that the value of $a_{i,j}$ is changed, each of these sets require the update of an element in the butterfly node, in addition to a single element in the horizontal node. Therefore, the expected number of updated elements is $\lfloor k/2\rfloor\cdot 1/2+2$.

In the worst case, consider the update of an element $a_{0,j}$, for $j\in[k]$. For each $j'\in [k]\setminus \{j\}$ such that $j'-j\le\lfloor k/2\rfloor\text{ (mod $k$)}$, $i(j')=i(j'-1)$, and therefore, $(0,j)\in B_{\ell_{0,j'}}$. For that reason, $\lfloor k/2\rfloor+1$ elements of the butterfly node need to be updated, in addition to a single element in the horizontal node, and the total is $\lfloor k/2\rfloor+2$.
\end{IEEEproof}

\section{Conclusions}
\label{sec:conclusions}

In this paper, we described a construction of repair-optimal MDS array codes, whose array elements are bits, and the operations are performed over GF($2$). Several problems are still open in this topic. First, it could be interesting to find out whether there exist repair optimal MDS codes with lower update measure. Second, a generalization of the construction to more parity nodes could be very useful. And finally, it would be important to know whether such codes exist whose number of rows is polynomial in the number of columns.

\section{Acknowledgments}
This work was done while Eyal En Gad was at HGST Research.
We would like to thank Yuval Cassuto for his help in reducing the worst-case update measure of the codes.

\bibliographystyle{IEEEtranS}
\bibliography{allbib}

\end{document}